\documentclass[12pt]{article}

\begin{document}
\begin{titlepage}

\author { VICTOR NOVOZHILOV \footnote{vnovo@vn9711.spb.edu} and
  YURI NOVOZHILOV \footnote{yunovo2002@yandex.ru} \\
 V.Fock Institute of Physics, St.Petersburg State University,\\
  Ulyanovskaya str. 1, 198504, St.Petersburg, Russia}
\title{\bf   CHIRAL PARAMETRIZATION OF QCD VECTOR FIELD IN SU(3)}
\date{}
\maketitle

\abstract {The chiral parametrization of gluons in $SU\left(
3\right)$ QCD is proposed extending an approach developed earlier
for $SU\left( 2\right)$ case. A color chiral field is introduced,
gluons are chirally rotated, and vector component of rotated
gluons is defined on condition that no new color variables
appeared with the chiral field. This condition associates such a
vector component with $SU\left( 3\right)/U\left( 2\right)$ coset
plus an  $U\left( 2\right)$ field. The topological action in
$SU\left(3\right)$  QCD is derived. It is expressed in terms of
axial vector component of  rotated gluons. The vector field in
$CP^2$ sector is studied in new variables of chiral
parametrization.}

\vskip 6pt

\end{titlepage}

\section{Introduction}

\vskip 9pt

Confinement still remains one of the most important problems of
QCD. The monopole condensation scenario for confinement
\cite{nambu,Hooft,Mandelstam,Polyakov} is considered as the most
probable way. Therefore, much efforts were undertaken last decade,
in order to find a proper parametrization of the gauge field in
pure QCD (without quarks), which could contain necessary
topological properties. The Faddeev soliton picture
\cite{Faddeev75} of QCD excitations was revived in the
Faddeev-Niemi knot model \cite{faniemi}, and two different
approaches to the parametrization of the QCD gauge field were
proposed \cite{faniemi1,faniemi07}. One of them was related to the
Cho decomposition of the QCD gauge field which included explicitly
the magnetic connection \cite{cho,cho1}.

However, restriction to the pure QCD raises the question, whether
such approximation is good. Due to the chiral anomaly
\cite{adler,BJ}, the color gauge sector and the color chiral
anomalous sector are parts of the total color space. An experience
of the Schwinger model and related phenomenological discussion of
4d case shows \cite{DJP} that an impact of anomaly is important.
The color chiral bosonization \cite{A+N} describes the anomalous
sector in terms of an effective action, and it was found that such
an action with a chromomagnetic vacuum background field admits
stable soliton-skyrmion solutions \cite{N+N,TMPh}. If we neglect
anomaly, i.e. if we neglect quark chiral phases then these
solitons disappear. The problem of color degrees of quarks and
gluons consists in their interrelation: are these degrees of
freedom independent, or taking into account quark color chiral
phases does not change the number of color degrees of freedom. The
case of SU(2) color \cite{novozhilov} shows that the quark color
field adds topological defects.

In this paper we consider $SU(3)$ color and extend to this case
the chiral parametrization of the QCD gauge field developed for
SU(2). Due to interdependence of chirally rotated gluons and
induced axial vector field, the chiral parametrization for $SU(3)$
involves fields in $CP^2$ space and $U(2)$. We consider background
field for gluons and quarks at one-loop level and calculate the
Wess-Zumino-Witten topological action for the color chiral field
expressed in terms of gluon variables.

In section 2 we introduce the left-right group, the chiral field
and anomaly. In section 3 we explain special features of chiral
parametrization in the case of $SU(3)$ and introduce $CP^2$, as
underlying color space. In section 4 we present the topological
action. Section 5 studies the $CP^2$ dynamics of the vector
component of the chiral parametrization. Section 6 discusses
results.

\section{Left-Right group $SU(3)_L\times SU(3)_R$ , chiral
field and the anomaly}

In this section we introduce Left-Right group $SU(3)_L\times
SU(3)_R$ , chiral field and the anomaly. Consider massless
fermions in external vector and axial vector fields $V_\mu ,A_\mu
$ and the Dirac operator

$$
\not D\left( V,A\right) =i\gamma ^\mu \left( \partial _\mu +V_\mu
+i\gamma _5A_\mu \right) \eqno(1)
$$
with antihermitean fields in algebra su(3), so that $V_\mu =-V_\mu
^ait_a $ ,$A_\mu =$ -$A_\mu ^ait_a$ , where $t_a,a=1,2...8$ are
generators. The chiral transformation of fermions is given by

$$
\psi _L^{\prime }=\xi _L\psi _L,\psi _R^{\prime }=\xi _R\psi
_R,\psi =\psi _L+\psi _R\eqno(2)
$$
where $\xi _L\left( x\right) $ and $\xi _R\left( x\right) $ are
local chiral phase factors of left and right quarks $\psi _L$ and
$\psi _R$ , represented by unitary matrices in defining
representations of left $SU\left( 3\right) _L $ and right
$SU\left( 3\right) _R$ subgroups of the chiral group $G_{LR}=$
$SU\left( 3\right) _L\times SU\left( 3\right) _R$ . For $\psi
_L=\frac 12\left( 1+\gamma _5\right) \psi ,\psi _R=\frac 12\left(
1-\gamma _5\right)
\psi $ , generators $t_{La}$ and $t_{Ra}$ of left and right subgroups of $%
G_{LR}$ can be written as $t_{La}=\frac 14\left( 1+\gamma
_5\right) \lambda _a,t_{Ra}=\frac 14\left( 1-\gamma _5\right)
\lambda _a,\left[ t_{La,}t_{Rb}\right] =0$ , where $\lambda
_a,a=1,2,..8$ are the Gell-Mann matrices. Then quark left and
right chiral phase factors $\xi _L,\xi _R$ arise from application
of operators $\hat \xi _L=\exp (-it_{La}\omega
_{La}),\hat \xi _R=\exp (-it_{Ra}\omega _{Ra})$ to left and right quarks $%
\psi _L$ and $\psi _R$. Vector gauge transformations $g\left(
x\right) $ are associated with $t_a=t_{La}+t_{Ra}=\lambda _a/2$ ,
i.e. $g\left( x\right) $ has properties of the product $\hat \xi
_L\left( x\right) \hat \xi _R\left(
x\right) $ of identical left and right rotations, $\omega _L=\omega _R$ $%
=\alpha $. The generator of purely chiral transformations
$g_5\left(
x\right) $ is $t_{5a}=\gamma _5\lambda _a/2=t_{La}-t_{Ra}$ ; thus, $%
g_5\left( x\right) $ has properties of $\hat \xi _L\left( x\right)
\hat \xi _R^{+}\left( x\right) $ for $\omega _L=\omega _R=\Theta $
. Infinitesimally, the Dirac operator is transformed according to
$$
\delta \not D=[i\frac 12\alpha _a\lambda _a,\not D]+\{i\frac
12\gamma _5\Theta _a\lambda _a,\not D\} \eqno(3)
$$

Commutation relations for $t_a,t_{5a}$ are given by

$$
\left[ t_a,t_b\right] =if_{abc}t_c,\left[ t_a,t_{5b}\right]
=if_{abc}t_{5c},\left[ t_{5a},t_{5b}\right] =if_{abc}t_c\eqno(4)
$$
where $f_{abc}$ are antisymmetrical structure constants of
$SU\left( 3\right) $ .

Instead of phases $\xi _L$ and $\xi _R$ one can work with the chiral field $%
U=\xi _R^{+}\xi _L$, which describes rotation of only left quark
leaving right quark in peace $\psi _L\rightarrow \psi _L^{\prime
}=U\psi _{L,}\psi \rightarrow \psi _R^{\prime }=\psi _R$ . The
same result can be obtained by the chiral transformation $\psi
_L\rightarrow \xi _L\psi _L,\psi _R\rightarrow \xi _R\psi _R$ ,
followed by a vector gauge transformation with a gauge function
$\xi _R^{+}$ . The usual chiral gauge choice is $\xi _R=\xi
_L^{+}$ , then the chiral field is taken as squared left chiral
phase:

$$
U=\xi _L^2=\exp i\Pi ,\Pi =\Pi _a\lambda _a\eqno(5)
$$
where we used the flavor notation $\Pi $. To describe $U$ one can
use the the Cartan basis with diagonal matrices $H_1,H_2$ and step
up/down operators
$$
E_{\pm 1}=\frac 1{2\sqrt{2}}\left( \lambda _1\pm i\lambda
_2\right) ,E_{\pm 2}=\frac 1{2\sqrt{2}}\left( \lambda _4\pm
i\lambda _5\right) ,E_{\pm 3}=\frac 1{2\sqrt{2}}\left( \lambda
_6\pm \pm i\lambda _7\right)
$$

For applications related to SO(3) monopoles and $SU(3)/SO(3)$
coset, it is convenient to consider $SU(3)$ in the $SO(3)$ basis
\cite{winternitz}. $SO(3)$ is the maximal subgroup of $SU(3)$. We
use two hermitian combinations: $\hat N\left( x\right) $ and $\hat
N^2\left( x\right) $ , where $\hat N\left(
x\right) $ is built on antisymmetric $\lambda $-matrices $\hat N=n_kO_k$ , $%
O_k=\left( \lambda _7,-\lambda _5,\lambda _2\right) $ ,
$n_kn_k=1,$ $N^3=N$ , while $\hat N^2$ contains only symmetric
$\lambda $'s. Then general $SU(3)$ -chiral field is
$$
U\left( \alpha ,\beta \right) =\exp i\Pi \left( \alpha ,\beta
\right) =\exp i\left( \hat N\alpha +(\frac 12\{\hat N^{\prime
},\hat N^{\prime \prime }\}-\frac 13trN^{\prime }N^{\prime \prime
})\beta \right), \eqno(6)
$$
 where $\hat N^{\prime },\hat N^{\prime \prime }$ depend on
$SO(3)$ unit
vectors $n_k^{\prime },n_l^{\prime \prime }$ . Three unit vectors plus $%
\alpha ,\beta $, give altogether 8 parameters of SU(3).

The chiral tranformation of fermions in the Dirac action is
equivalent to the following change of the Dirac operator
$$
\bar \psi ^{\prime }\not D\left( V,A\right) \psi ^{\prime } =\bar
\psi D\left( V^U,A^U\right) \psi $$
$$
V^U=\frac 12[U^{+}(\partial +V+A)U+(V-A)],A^U=\frac
12[U^{+}(\partial +V+A)U-(V-A)]\eqno(7)$$

Consider now the fermionic path integral $Z_\psi \left[ V,A\right]
=\int d\mu \exp i\int dx\overline{\psi }\not D\left( V,A\right)
\psi $ . Our starting point is such $Z_\psi ^{\prime }$ , where
fermion chiral phases are extracted from fermions by
transformation $\psi \rightarrow \psi ^{\prime }$ in the Dirac
action. A transformed fermionic path integral $Z_\psi ^{\prime } $
is equal to an original path integral as a functional of
transformed fields $Z_\psi ^{\prime }\left[ V,A\right] =Z_\psi
\left[ V^U,A^U\right] $ . Thus, we are going to calculate the path
integral
$$
Z_\psi ^{\prime }\left[ V,A\right] =Z_\psi \left[ V^U,A^U\right]
=\int d\mu \exp i\int dx\overline{\psi }\not D\left(
V^U,A^U\right) \psi , \eqno(8)
$$
where fields are chirally rotated: $V,A\rightarrow $ $V^U,A^U$ .
The path integral $Z_\psi $ is invariant under vector gauge
transformations of fermions, but undergoes changes under chiral
transformations, because of non-invariance of the fermionic
measure $d\bar \psi d\psi $ \cite{fujikawa}: chiral
transformations are anomalous. The chiral anomaly $\mathcal{A}$ is
defined by an infinitesimal change of $\ln Z_\psi $ due to an
infinitesimal chiral transformation $\delta g_5=i\theta _a\tau
_a\equiv \Theta $.

We put $g_5(s)=\exp \gamma _5\Theta s$ and write the anomaly $\mathit{A}%
\left( x,\Theta \right) $ at a chiral angle $\Theta $
$$
\mathit{A}\left( x,\Theta \right) =\frac 1i\frac{\delta \ln Z_\psi
\left( \exp \Theta s\right) }{\delta s}_{s=1}\eqno(9)
$$
In flavor bosonization the main question is, what is an effective
action for the chiral field as a new independent variable. The
usual way [12] to calculate such an effective chiral action
$W_{eff}$ is to find the anomaly and integrate it over $s$ up to
$g_5=\exp \gamma _5\Theta $
$$
W_{eff}=-\int d^4x\int_0^1ds\mathit{A}\left( x;s\Theta \right)
\Theta \left( x\right) =\int dxL_{eff}-W_{WZW},\eqno(10)
$$
where the Wess-Zumino-Witten term $W_{WZW}$ describes topological
properties of the chiral field $U$ and is represented by a
five-dimensional integral with x$_5=s$. In general, it cannot be
expressed as a 4-dimensional integral over density $L_{WZW}$. The
first term in $W_{eff}$ is of non-topological nature. It is the
analogue of $W_{WZW}$ for color that we are interested in.
However, in the case of color, the vector field $V_\mu $ is a
dynamical one describing gluons, a dynamical axial vector field
$A_{\mu }$ does not exist and the chiral field $U$ cannot be
considered as an independent additional variable.

\section{Chiral parametrization of the QCD vector field}

In this section we apply the scheme sketched in the previous
section to the case of the QCD vector field, which from now on is
denoted by $V_\mu $ . It is the field that enters the Dirac
operator for quarks, and we should consider it as a background
field. In absence of a dynamical axial vector color field, $A^U$
denotes an axial component of the chirally rotated gauge field

$$
V^U=\frac 12[U^{+}(\partial +V)U+V],A^U=\frac 12[U^{+}(\partial +V)U-V)]%
\eqno(11)
$$
Thus, our initial setting for gluons plus quarks can be expressed
by the following path integral
$$
Z[V,U]=\int d\mu _Q\exp iI_{eff}\left( V+Q\right) Z_\psi \left[
V^U,A^U\right] \eqno(12)
$$
where $Q$ is a quantum field and $I_{eff}$ is the QCD effective
action including the Faddeev-Popov ghosts and gauge fixing. In
this paper, we are not going to consider an integration over $Q$ ,
but but we shall integrate over quarks in order to get the
Wess-Zumino-Witten topological action in terms of the chiral field
$U$ . We work with an initial gauge field $V_\mu $ and two fields
arising in its chiral transformation $U$ , namely, a gauge field
$V_\mu ^U$ and an axial vector field $A_\mu ^U$ .However, by
introducing the quark color chiral field $U$ we arrive at a system
with too many degrees of freedom. A consistent parametrization of
gluons and the chiral field within Left-Right color group becomes
a special task. To eliminate superfluous variables, one should
consider a relation between gluonic fields $V_\mu $ and chirally
transformed field $V_\mu ^U$ and find how different parts of these
fields can be made from the same material, so that chiral field
variables are either fixed, or fully incorporated into gauge field
variables. This is a key point in our approach. When this task is
accomplished, one can integrate chiral color anomaly and get a
topological term, an analogue of the Wess-Zumino-Witten action .

Two simple expressions exist for $V^U\pm A^U$ , namely
$$
V_\mu ^U+A_\mu ^U=U^{+}(\partial _\mu +V_\mu )U,V_\mu ^U-A_\mu ^U=V_\mu %
\eqno(13)
$$
which mean that if the chiral field is considered as a regular
gauge transformation, then ($V^U\pm A^U)_\mu $ - combinations
should have the same field strengths $\left( V^U\pm A^U\right)
_{\mu \nu }$ . With topological $U$ these combinations can
describe different situations.

In order to find a partial parametrization of gluonic field in terms of
chiral parameters, let us at first define a combination of the gauge field $%
V_\mu ^U$ and the chiral field $U$, which is invariant under chiral
transformation :
$$
(V_\mu ^U)^U=V_\mu ^U\eqno(14)
$$
A gauge field with this property we denote $V_\mu ^\Omega $ . It will
contain $U$-variables. An axial vector field $A_\mu ^\Omega $ calculated
with $V_\mu ^\Omega $ instead of $V_\mu $ is absent: $A_\mu ^\Omega \equiv
\frac 12[U^{+}(\partial _\mu +V_\mu ^\Omega )U-V_\mu ^\Omega ]=0$ . It
follows also that
$$
U^2=\exp i\zeta ,\partial _\mu \zeta =0\eqno(15)
$$
independently of the chiral color group. The invariance relation cannot be
true in the whole space of chiral color, but it can be satisfied in a region
with restricted number of variables. Such a region can be found by studying
the ($V_\mu ^U\rightarrow V_\mu )$ - determinant.

Consider relation between 8$\times 8$ matrices of gluonic field $V_\mu $ and
chirally rotated field $V_\mu ^U\left( x\right) $ in adjoint representation
$$
(V_\mu ^U)_{ab}=\frac 12\left( 1+R\left( U\right) \right)
_{ab}V_{\mu b}+\frac ig\frac 12(U\partial _\mu
U^{+})_{ab},R_{ab}\left( U\right) =\frac 12tr\left( \lambda
_aU\lambda _bU^{+}\right)\eqno(16)
$$
where the chiral field $U=\xi _L^2=\exp i\Theta ,\Theta =\lambda _a\Theta
_a, $ is defined in the chiral gauge $\xi _L=\xi _R^{+}$ . Here $\lambda
_a,a=1,2..8$ , are the Gell-Mann matrices.

We write $U$ in flavor-like notation
$$
U=\exp i\Pi ,\Pi =\lambda _a\Pi _a\eqno(17)
$$
It is also convenient to write $\Pi $ in the SO$\left( 3\right) $ basis (see Section 2).

 In order to calculate the determinant $\det \frac 12\left( 1+R\left( U\right) \right) $,
it is sufficient to consider the case $N^{\prime }=N^{\prime \prime }=N$ ,
when
$$
U\rightarrow U\left( \hat N,\alpha ,\beta \right) =\exp \left(
-\frac 23i\beta \right) \left[ 1+i\hat Ne^{i\beta }\sin \alpha
+\hat N^2\left( e^{i\beta }\cos \alpha -1\right) \right]\eqno(18)
$$
When eigenvalues of $\hat N$ are placed as diag$\left( 1,-1,0\right) $, we
see that $\alpha =\Theta _3,\beta =\Theta _8\sqrt{3}$ and

$$
\det \frac 12\left( 1+R\left( U\right) \right) =\frac 12\left(
1+\cos 2\alpha \right) \frac 12\left( 1+\cos \left( \alpha +\beta
\right) \right) \frac 12\left( 1+\cos \left( \alpha -\beta \right)
\right)\eqno(19)
$$
This result can be easily checked in (isospin I$_3$, hypercharge
Y) basis, where the diagonal elements of R$\left( U\right) =\exp
i\left( I_32\alpha +Y2\beta \right) $ , as of the adjoint
representation of U, are nothing, but those of octet: pions (I=1,
Y=0), K-mesons (I=1/2, Y=1/2), \=K-mesons (I==1/2, Y=-1/2),
$\sigma -$meson (I=0,Y=0).

This determinant is invariant under reflection $\left( \alpha ,\beta \right)
\rightarrow \left( -\alpha ,-\beta \right) $ . It disappears for values $%
\left( \alpha ,\beta \right) $ equal to $\left( \pi /2,0\right) ,$ $\left(
\pi /2,\pm \pi /2\right) $ and $\left( 0,\pi \right) $ characterizing
singularity surfaces in chiral color space (i.e. $\gamma _5\Theta $ ). The
first of these sets, $\left( \pi /2,0\right) $ , corresponds to $SO\left(
3\right) $ subgroup with one zero factor in $\det $ , when $U\left( N,\pi
/2,0\right) =\exp i\hat N\alpha =1+i\hat N-\hat N^2$ .

For $SU\left( 3\right) $ we are interested in two coinciding zero
factors of $\bf{\det }$ related to two simple roots of $SU\left(
3\right) $ . Together with a singularity set related to a complex
root, we can define three color chiral fields $U\left( \hat
N,\alpha ,\beta \right) $ for $SU\left( 3\right) $ . For the pair
$\left( \alpha ,\beta \right) =\left( 0,\pi \right) $ we get the
chiral field

$$
U\left( \hat N,0.\pi \right) =(1-2\hat N^2)\exp (-i\frac 13\pi
)=m_1\exp (-i\frac 13\pi )\eqno(20)
$$

For $\left( \alpha ,\beta \right) =\left( \pi /2,\pi /2\right) $ we have

$$
U\left( \hat N,\pi /2,\pi /2\right) =\left( 1-N-N^2\right) \exp
(-i\frac \pi 3)=m_2\exp (-i\frac \pi 3)  \eqno(21)
$$
We denote $U\left( \hat N,\pi /2,-\pi /2\right) =m_3\exp \left(
-i\pi /3\right) $. In all these cases $m_k$ are normalized
hermitian 3$\times 3$ matrices in color space : $m^2=1$. A product
of two $m$ 's is equal to the
third $m$ up to a constant phase. Their diagonal forms are $m_1^0=$ diag$%
\left( -1,-1,1\right) ,m_2^0=$diag$\left( -1,1,1\right) $ and $m_3^0=$diag$%
\left( 1,-1,1\right) $. The same diagonal forms correspond to
skyrmion-type embeddings of SU(2) in SU(3), when $\Pi =\lambda
_k\Pi _k+E_{33},E_{33}=$ diag$\left( 0,0,1\right) $. In the
general case, matrices $m_k$ are given by
$$
m_k=Sm_k^0S^{+}\eqno(22)
$$
where $S$ is a unitary SU(3) transformation common for all $k=1,2,3.$ Then $%
S $ is defined up to right multiplication by a diagonal unitary matrix, i.e $%
S$ is in the coset $SU(3)/U(1)^2$ and depends on six parameters
associated with nondiagonal $\lambda ^{\prime }$s. When each $m$
is considered separately, so that $m_k=S\left( k\right)
m_k^0S^{+}\left( k\right) $, then $S\left( k\right) $ is in the
coset $SU(3)/U(2)$, and different $m_k$ do not commute. Unlike
SU(2) unit matrix, the SU(3) matrices $m_k$ are not traceless.
Traceless diagonal matrices $m_k^0-1/3,k=1,2,3,$ are in the
following relation with roots $r\left( k\right) $:
$$
m_k^0-1/3=\pm D^{\left( k\right) }, \eqno(23)
$$
where $D^{\left( k\right) }$ gives $SU(3)$ magnetic monopole
embedding along roots $r\left( k\right) $.

Chirally invariant vector field $V_\mu ^\Omega $. Behavior of the
($V_\mu ^U\rightarrow V_\mu )$ determinant $\det \frac 12\left(
1+R\right) $ shows, how to restrict chiral color space in defining
the chiral field $U$, in order to construct the field $V_\mu
^\Omega $. Gauge invariant chiral
regions $\Omega $ are given by sets $\left( \alpha ,\beta \right) $ = $%
\left( 0,\pi \right),\left( \pi /2,\pm \pi /2\right) $ , where the
chiral field $U=m_k$ is represented correspondingly by one of
3$\times 3$ unit matrices $m_k,k=1,2,3$ up to a constant phase.

In the case of $SU\left( 2\right) $ color, the chiral
parametrization of gluons \cite{novozhilov} defines the vector
component $V_\mu ^\Omega =\hat nC_\mu +\frac 12\hat n\partial _\mu
\hat n$ in terms of unit vector $\hat n$ $=\tau _kn_k$ belonging
to the chiral field and which is covariantly constant: $D_\mu
\left( V_\mu ^\Omega \right) \hat n=0$ . The field $V_\mu ^\Omega
$ was introduced first in the decomposition of gluons in order to
include topological structures into gluonic decomposition
\cite{cho},\cite {faniemi} and exploit an idea of the abelian
dominance. It was done without
recourse to the chiral anomaly. Similar approaches to $SU\left( 3\right) $ %
\cite{bolokhov} ,\cite{shabanov} are based on the Cartan algebra $%
H_k,k=1,2,3,$ and lead to $SU\left( 3\right) /U\left( 1\right) ^2$
submanifold $S$ with two independent vectors $n_k=SH_kS^{-1}$ and two
independent $V_\mu ^\Omega $ -like fields.

The chiral parametrization changes the situation. Let us formulate
first the result and then discuss the difference with the case,
when the anomaly is neglected. The chiral parametrization of
gluons in $SU\left( 3\right) $ QCD contains only one unit
$SU\left( 3\right) $ matrix $m,m^2=1,tr m=1$, defining a chiral
field $U=m$ , a direction in $SU\left( 3\right) $ space. The
chirally invariant vector field $V_\mu ^\Omega $ is given by

$$
V_\mu ^\Omega = C_\mu \left( m-1/3\right) + G_\mu +\frac
12m\partial_\mu m,\eqno(24)
$$
where $m$ denotes one of three unit matrices $m_k$ and $G_\mu$ is
 a $U\left(2\right)$ component of $V_\mu^\Omega$,
  so that $\left[m,G_\mu\right]=0$. It can be considered as
a result of chiral rotation $U=m$ applied to a pair $\left(V_\mu,
A_\mu\right) =\left(C_\mu m +G_\mu  ,\frac 12m\partial _\mu
m\right)$.
 The chirally rotated vector component reproduces itself,
$$
(V_\mu ^\Omega )^U=V_\mu ^\Omega ,V_\mu ^\Omega =\frac 12\left(
UV_\mu ^\Omega U^{+}+V_\mu ^\Omega +U\partial _\mu U^{+}\right)
,U=m\eqno(25)
$$
while $A_\mu ^U$ , as an axial complement of $V_\mu ^\Omega $,
disappears
$$
A_\mu ^U=\frac 12UD_\mu \left( V^\Omega \right) U^{+}=0\eqno(26)
$$
The matrix $m$ is covariantly constant: $D_\mu \left( V^\Omega
\right) m=0$.

In general, the chiral matrix $\Pi $ can be written as a sum over
two independent roots, or two $m_k$'s, like ($am_1+bm_2)$ instead
of single $m.$ However, the invariance condition can be satisfied
only with $a=\pm b=\pi /2$ , or $U=m_3$. Therefore, $U$ and
$m\partial m$ term depend always only on one matrix $m.$ It can be
interpreted that the field $V_\mu ^\Omega $ asymptotically might
describe only one-monopole configurations.

Thus, the chiral anomaly imposes important restrictions on chiral
field $U$ and subspace of $SU\left( 3\right) $, which can be used
for chiral parametrization of gluonic field $V_\mu $. Basic matrix
$m,$ which is a counterpart of traceless $SU\left( 2\right) $ unit
matrix $\hat n$ , cannot be traceless, because it comes from an
unitary matrix. Though we have choice of three possibilities $m_k$
, the $SU\left( 3\right) $ QCD uses only one. We shall take it as
$m^0=diag\left( 1,1,-1\right) =2Y+1/3$ , where $Y$ denotes the
color hypercharge. This corresponds to the chiral field $U=\exp
iY\pi $. Then $Sm^0S^{-1}$ is the orbit of $SU\left( 3\right) $
through $Y$. The $SU\left( 3\right) $ matrix $S$ is defined up to right multiplication $%
S\rightarrow Sg$ , $g\in U\left( 2\right) $, where $U\left(
2\right) $ is built on $\lambda _k,k=1,2,3,8.,$ so that
$CP^2=SU\left( 3\right) /U\left( 2\right) $ , while $S$ includes
$\lambda _A,A=4,5,6,7.$ Unit matrix $m$
commutes with $\lambda _k$ and anticommutes with $\lambda _A$. Matrix $%
\tilde m$ can be identified with $\lambda _3$.

Induced axial vector field $A_\mu ^U$. Consider basic relation
$V_\mu ^U-A_\mu ^U=V_\mu $. It follows from the structure of
chirally rotated fields $V_\mu ^U$ and A$_\mu ^U$ that any field
$B_\mu $ commuting with matrix $U=m$ contributes only to rotated
vector field $V_\mu ^U$ , while a field $X_\mu $ anticommuting
with $m$ contributes only to the induced axial vector field $A_\mu
^U$ , and this property does not depend on a particular color
group. Thus, the latter set of $\lambda $'s can be used to build a
contribution $X_\mu $ to field $A_\mu ^U$%
$$
X_\mu =S\lambda _ax_{a\mu }S^{+},a=4,6,6,7,\eqno(27)
$$
with the property $\{X_\mu ,m\}=0$. Another example of matrices
anticommuting with $m$ provide derivatives $\partial _\mu m$ and
$m\partial
_\mu m$ , which we use to construct a contribution $Y_\mu $ to $A_\mu ^U$%
$$
Y_\mu =\varphi \partial _\mu m+i\chi m\partial _\mu m\eqno(28)
$$
where $\varphi $ and $\chi $ are colorless functions. Such terms
exist in the two-color decomposition \cite{faniemi}. The
decomposition of gluons $V_\mu $ is given by
$$
V_\mu =V_\mu ^U-A_\mu ^U=V_\mu ^\Omega -\hat{\mathit{A}_\mu}
,\hat{\mathit{A}_\mu} =X_\mu +Y_\mu  \eqno(29)
$$
It is easy to check that an axial field $\hat{\mathit{A}_\mu}$
arises in chiral rotation $U=m$ from $V_\mu =V_\mu ^\Omega
-\hat{\mathit{A}_\mu} $. If before chiral rotation $U=im$ the
quark path integral depends on the gauge field only, $Z_\psi
\left[ V_\mu ,0\right] =Z_\psi \left[ V_\mu ^\Omega
-\hat{\mathit{A}_\mu} ,0\right] $ , then after this rotation we
get $Z_\psi \left[ V_\mu ^\Omega ,\hat{\mathit{A}_\mu} \right] $.

Thus, as it follows from expressions for the QCD vector field ('gluons') $%
V_\mu $ there are two distinct sectors in the chiral decomposition
of $V_\mu =V_\mu ^\Omega -\hat{\mathit{A}_\mu}$:

(a) the $CP^2$-sector with the dynamical abelian field $C_\mu $.
The space $CP^2$ enters the scene with the chiral field
$U=m=Sm_0S^{+}$, as the $SU(3)$ orbit through $m_0$. The matrix
$m$ is the main element, which defines the vector field in the
sector $\left( V_\mu ^\Omega \right) _{CP^2}=mC_\mu +\frac
12m\partial _\mu m$, including the direction of an abelian field
$C_\mu $ in the $SU(3)$ space. The axial component
$\hat{\mathit{A}_\mu} $ anticommutes with $m$.

(b) an $U\left( 2\right)$ sector with the  field $\left( V_\mu^
\Omega \right)_{U\left( 2\right)}= G_\mu$; the chiral field $U=m$
commutes with $G_\mu$.

\section{Topological action}

The effective action $W_{eff}$
$$
W_{eff}=W_{WZW}+W_{an}=-i\ln \{Z_\psi \left[ V_\mu ^\Omega
,\hat{\mathit{A}}_\mu \right] Z_\psi ^{-1}\left[ V_\mu ^\Omega
-\hat{\mathit{A}}_\mu ,0\right] \eqno(30)
$$
describes the quark color chiral contribution to QCD dynamics and
contains the topological term $W_{WZW}$, which is an analogue of
the chirally gauged Wess-Zumino-Witten action in flavor physics
\cite{WZ}, \cite{witten}, as well as a non-topological action
$W_{an}$. Non-topological term $W_{an}$ has the same structure in
$SU(3)$ as in $SU(2)$.

The gauged Wess-Zumino-Witten term $W_{WZW}$  adapted for the
general color case, (i.e. no dynamical axial vector, $A_\mu =0$,
and $U$ belongs to $SU\left( 3\right) $), is given in the
Minkowski space by the five-dimensional integral (see also
\cite{andrianov} )
$$
W_{WZW}=\frac i{96\pi ^2}\int d^5x\varepsilon _{\mu \nu \sigma
\lambda \rho }tr[(j_\mu ^{-}+j_\mu ^{+})V_{\nu \sigma }V_{\lambda
\rho }+
$$

$$
 \frac 12\left( j_\mu ^{-}V_{\nu \sigma }U_sV_{\lambda \rho
}U_s^{-1}+j_\mu ^{+}V_{\nu \sigma }U_s^{-1}V_{\lambda \rho
}U_s\right) -iV_{\mu \nu }(j_\sigma ^{-}j_\lambda ^{-}j_\rho
^{-}+j_\sigma ^{+}j_\lambda ^{+}j_\rho ^{+})
$$
$$
-\frac 25j_\mu ^{-}j_\nu ^{-}j_\sigma ^{-}j_\lambda ^{-}j_\rho
^{-}]\eqno(31)
$$
where the following notations are used
$$
j_\mu ^{-}=D_\mu U_sU_s^{-1},j_\mu ^{+}=U_s^{-1}D_\mu U_s,U_s=\exp
s\Theta
$$
$$
D_\mu U_s=\partial _\mu U_s+[V_\mu ,U_s],x_5=s
$$
and the convention $\mu ,\nu ,...=1,2,3,4,5;L_5=R_5=0$ implied.
The fifth integration here is an $s-$integration over chiral
anomaly. An additional integration arises from the relation $\ln
Z_\psi \left[ V,0\right] -\ln Z_\psi \left[ V^U,A^U\right] =-\int
ds\partial _s\ln Z_\psi \left[ U\left(s\right) \right] $ with
boundary points $s=0,$ $U\left( 0\right) =1$ , and $%
s=1$ ,$U\left( 1\right) =U$ . For the case under consideration, when $V_\mu
^U=V_\mu ^\Omega ,A_\mu ^U=\hat{\mathit{A}_\mu} ,V_\mu =V_\mu ^\Omega -\hat{\mathit{A}_\mu} $ , we put $U\left( s\right) =\exp \{ims\pi /2\}$ using relations $%
D_\nu ^\Omega m=0$ and $m\hat{\mathit{A}_\mu}
+\hat{\mathit{A}_\mu} m=0$ . An integration over $s$ is simple, so
that in the case of color bosonization there exists a topological Lagrangian corresponding to $W_{WZW}$%

$$
L_{WZW}=\frac 1{32\pi }\varepsilon _{\nu \sigma \lambda \rho }trm\{\frac
12V_{\nu \sigma }^\Omega V_{\lambda \rho }^\Omega -V_{\nu \sigma }^\Omega (%
\hat{\mathit{A}}_{\lambda \rho }+[\hat{\mathit{A}}_\lambda
,\hat{\mathit{A}}_\rho ])-\left[
\hat{\mathit{A}}_\nu ,\hat{\mathit{A}}_\sigma  \right] \hat{\mathit{A}}_{\lambda \rho }]\}\\
$$
$$
\hat{\mathit{A}}_{\nu \sigma }=D_\nu ^\Omega
\hat{\mathit{A}}_\sigma -D_\sigma ^\Omega
\hat{\mathit{A}}_\nu ,D_\nu ^\Omega =\partial _\nu +\left[ V_\nu ^\Omega ,*\right] %
\eqno(32)
$$
where $D^\Omega $ contains the field $V_\mu ^\Omega $ , while
$V_{\mu \nu }^\Omega $ is the field strength of $V_\mu ^\Omega $.
The Wess-Zumino-Witten part does not contain terms with four fields\textit{\ }$%
\hat{\mathit{A}} $.

\section{Vector component $V_\mu ^\Omega $ in $CP^2$ sector}

We are interested in variables appearing in $CP^2$ sector of the
chiral decomposition of gluons $V_\mu =V_\mu ^\Omega
-\hat{\mathit{A}}_\mu $. To this end, we consider the vector
component $V_\mu ^\Omega $ only in $CP^2$ -sector, i.e. we neglect
the $U(2)$ field $ G_\mu $ and the axial part
$\hat{\mathit{A}}_\mu $. Then
$$
\Gamma _\mu =(V_\mu ^\Omega )_{CP2}=C_\mu \hat m^{\prime }+\frac
12\hat m\partial _\mu \hat m\eqno(33)
$$
where $\hat m^{\prime }=m-1/3$. The field strength $\Gamma _{\mu
\nu }^{}$ is expressed in terms of basic unit $SU\left( 3\right) $
matrix $\hat m$
$$
\Gamma _{\mu \nu }^{}=C_{\mu \nu }\hat m^{\prime }+\frac 14\left[
\partial _\mu \hat m,\partial _\nu \hat m\right] \eqno(34)
$$
The matrix $\hat m$ is normalized according to
$$
\hat m^2=1,tr\hat m=1\eqno(35)
$$
and can be considered as a unitary transform of a constant diagonal matrix $%
\hat m_0=diag\left( 1,1,-1\right) $%
$$
\hat m=S\hat m_0S^{-1},SS^{+}=1\eqno(36)
$$
The unitary matrix $S$ is defined up to right U$\left( 2\right) $
transformations $g:S\rightarrow Sg,$ built on color analogues of isospin $I_C
$ and hypercharge $Y_C$ generators.
$$
g=\exp i\left( \lambda _k\beta _k+m_0\phi \right) \equiv g_2\exp
im_0\phi ,gm_0g^{-1}=m_0\eqno(37)
$$
Then $S$ can be written as a function of four parameters associated with
matrices $\lambda _4,\lambda _5,\lambda _6,\lambda _7$ and described by a
hermitian matrix $\alpha $ and an U$\left( 2\right) $- scalar $\omega $
$$
S=\exp (i\hat \alpha \omega ),tr\hat \alpha ^2=2\eqno(38)
$$
Explicitly

$$
\hat \alpha =\hat \xi + \hat \xi^{+}
$$
$$
\hat \xi =\frac 12 \left(\lambda_4+i\lambda_5\right)\xi_1 +\frac
12 \left(\lambda_6+i\lambda_7\right)\xi_2 \eqno(39)
$$

Here isospinor components $\xi _1,\xi _2$ with quantum numbers of
color $K-$ fields $I_C=\pm \frac 12,Y_C=1$ are normalized
according to $\xi _1^{+}\xi _1+\xi _2^{+}\xi _2=1$. Other
components of color octet ($\pi $ and $\eta )$ can be found in
$\alpha ^2$.

Under $m_0$ - preserving transformation $g$ the matrix $\alpha $ and
isospinor $\xi =\frac 12\left( 1+m_0\right) \alpha $ transform according to
$$
\alpha ^{\prime }=g\alpha g^{-1},\frac 12\left( 1+m_0\right)
\alpha ^{\prime }=\xi ^{\prime }=g_2\xi \exp im_0\phi \eqno(40)
$$
It is easy to verify that $\hat \alpha =\hat \alpha ^3$ , so that $\hat
\alpha $ -eigenvalues are $\pm 1,0$ . Thus, $S$ can be written as
$$
S=1+i\hat \alpha \sin \omega +\hat \alpha ^2(\cos \omega
-1)\eqno(41)
$$
Now, $P_{\pm }=\frac 12\left( \hat \alpha ^2\pm \hat \alpha \right) $ and $%
P_0=\left( 1-\hat \alpha ^2\right) $ are projection operators on
eigenvalues $\pm 1,0$. We can build an isovector $n_k$ from
isospinor $\xi $ with components $\xi _a=\left( \alpha \right)
_{ka},a=1,2$ and its conjugated
isospinor $\xi ^{+}$ with components $\xi _k^{+}=\left( \alpha \right) _{3a}$%
$$
n_k=\xi ^{+}\tau _k\xi \eqno(42)
$$
while $\left( \alpha ^2\right) _{ab}=\xi _a\xi _b^{+}$ and $\left(
\alpha ^2\right) _{33}=\xi ^{+}\xi =\xi _a^{+}\xi _a=1$ . An
$SU\left( 2\right) $ unit vector $\hat n$ is an useful collective
variable entering the knot model \cite{Faddeev75}. Isospinors $\xi
$ , $\xi ^{+}$ ,as members of octet, correspond to quark-antiquark
variables, while isovector $\hat n$ , as a bilinear combination of
$\xi ^{+}$,$\xi $ , may represent four-quark degrees of freedom,
though it belongs also to the same octet.

A relation between 3$\times 3$ matrix $\alpha $ , isospinor $\xi $ , 2$%
\times 2$ unit matrix $\hat n=\tau _kn_k$ and $3\times 3$ matrix $\mathbf{n}$
is given by
$$
\hat \xi =\frac 12\left( 1+m_0\right) \hat \alpha \\
$$
$$
\hat \alpha ^2=\frac 12\left( 1+\mathbf{n}\right) \ =(\hat
n+1)/2\eqno(43)
$$
It means also that in isospin subgroup a matrix $\hat \alpha ^2$
acts as a projector on eigenvalue $\hat n^{\prime }=+1$ , while in
$SU\left( 3\right) $
it is a projector on eigenvalues $\alpha ^{\prime }=\pm 1$. The matrices $%
\lambda _4,\lambda _5,\lambda _6,\lambda _7$ anticommute with
$m_0$, therefore
$$
\hat m=Sm_0S^{-1}=Um_0,U=S^2\eqno(44)
$$
Thus, the vector component $V_\mu ^\Omega $ of the gluonic field can be
expressed either in terms of 3$\times 3$ hermitian matrix $\alpha $ and an $SU%
\left( 3\right) $ scalar $\omega $, or in terms of isospin
doublets $\hat \xi $ and $\hat \xi ^{+}$ and $\omega $.

We are interested in composite variables based on elementary $\hat \xi $ and
$\hat \xi ^{+}$. Then it is sufficient to consider the case of constant $%
\omega $. We have for $\omega =\pi /4$, when $\left( U\right)
_{33}=0$, the following relations

$$
U\partial _\mu U^{+}E_{+} =-i\left( \eta _\mu +\eta_\mu
^{+}\right) +b_\mu \left( 1+\hat \xi \hat \xi ^{+}\right) +B_\mu
U\partial _\mu U^{+}E_{-} =b_\mu  \eqno(45)
$$
where$\ b_\mu $ and $B_\mu $ are isoscalar and $2\times 2$ matrix currents
$$
b_\mu =\left( \xi ^{+}\partial _\mu \xi \right) =-\left( \partial
_\mu \xi ^{+}\xi \right) ,B_\mu =\hat \xi \partial _\mu \hat \xi
^{+}-\partial _\mu \hat \xi \hat \xi\ \eqno(46)
$$
$\eta_\mu ,\eta_\mu ^{+}$ are gauge invariant isospinors
$$
\eta_\mu =\partial _\mu \xi -b_\mu \xi ,\eta_\mu ^{+}=\partial
_\mu \xi ^{+}+b_\mu \xi ^{+}\ \eqno(47)
$$
and $E_{\pm }$ are projectors $E_{\pm }=\frac 12\left( 1\pm
m_0\right)$. In the $SU(2)$ color case, the vector $b_\mu $ is
called magnetic potential.

The field strength $\Gamma _{\mu \nu }^{}$ in new variables $\hat \xi ,\hat
\xi ^{+}$ is given by
$$
\Gamma _{\mu \nu }^{}=C_{\mu \nu }(m-1/3)-\frac 14\left[ U\partial _\mu
U^{+},U\partial _\nu U^{+}\right] =\Gamma _{\mu \nu }^{+}+\Gamma _{\mu \nu
}^{-}+\Gamma _{\mu \nu }^\xi \\
$$
$$
\Gamma _{\mu \nu }^{+}=C_{\mu \nu }\left( 1-\hat \xi \hat \xi ^{+}\right)
-\frac 14h_{\mu \nu }\hat \xi \hat \xi ^{+}-\frac i4\left( \partial _\mu
\left( \hat \xi \hat \xi ^{+}\right) b_\nu -\partial _\nu \left( \hat \xi
\hat \xi ^{+}\right) b_\mu \right) \\
$$
$$
\Gamma _{\mu \nu }^{-} =-\frac 14h_{\mu \nu } \\
\Gamma _{\mu \nu }^\xi =-\frac i4\{h_{\mu \nu }\left( \hat \xi
-\hat \xi \right) -b_\mu \partial _\nu (\hat \xi -\hat \xi
^{+})+b_\nu \partial _\mu (\hat \xi -\hat \xi ^{+})\}\eqno(48)
$$

It follows that the field strength $\Gamma _{\mu \nu }^{}$ describes two
interacting systems. One of them is characterized be color isospin matrix
variables $\hat \xi ,\hat \xi ^{+}$ with density $\rho =\hat \xi \hat \xi
^{+}$ , the second one is characterized by hypercharge like potential $b_\mu
$ and field strength $h_{\mu \nu }$ . The latter variable is directly
related to an important quantity $tr\left( \hat m\left[ \partial _\mu \hat
m,\partial _\nu \hat m\right] \right) $ , which for $\omega =const$ reduces
to
$$
tr\left( \hat m\left[ \partial _\mu \hat m,\partial _\nu \hat
m\right] \right) =-h_{\mu \nu }\eqno(49)
$$
An expression of $h_{\mu \nu }$ in terms of the magnetic potential $b_\nu $
involves a supercurrent $\Sigma _{\mu \nu }$%
$$
h_{\mu \nu }=\partial _\mu b_\nu -\partial _\nu b_\mu -\Sigma _{\mu \nu }%
\eqno(50)
$$
where $\Sigma _{\mu \nu }=\left( \xi ^{+}\left[ \partial _\mu
,\partial _\nu \right] \xi \right) $. The quantity $h_{\mu \nu }$
can be associated with magnetic field strength. In this picture of
two interacting systems $\left( \hat \xi ,\hat \xi ^{+}\right) $
and $b_\mu $, kinetic terms in the Yang-Mills Lagrangian will be
quadratic.

The WZW Lagrangian for $\Gamma _\mu ^{}$ in new variables for the case $%
\omega =const$ is given by
$$
L_{WZW}=\frac 1{64\pi }\varepsilon _{\nu \sigma \lambda \rho }tr\{m\Gamma
_{\nu \sigma }^{}\Gamma _{\lambda \rho }^{}\}= \\
$$
$$
=\frac 1{64\pi }\int d^4x\varepsilon _{\nu \sigma \lambda \rho
}\{\frac 83C_{\mu \nu }C_{\lambda \sigma }-\frac 12C_{\mu \nu
}h_{\lambda \sigma }+h_{\mu \nu }h_{\lambda \sigma }w\left( \omega
\right) \eqno(51)
$$
where $w$ is a known functions of $\omega $.

Totally antisymmetric expression the action $W_{WZW}$ is reproducing itself
in partial integration, when
$$
\int d^4x\partial _\mu \varepsilon _{\mu \nu \lambda \sigma }tr\{\partial
_\nu m\partial _\lambda m\partial _\sigma m\}=0
$$
This leads to conservation of
$$
Q_m=\int d^3x\varepsilon _{ijk}tr\{\partial _im\partial _jm\partial _km\}%
\eqno(52)
$$
In general, in a similar manner the action $W_{WZW}\left( V_\mu
^\Omega \right) $ tells what kind of charges can be introduced in
$V_\mu ^\Omega $ -sector and what combination of charges is
conserved.

\section{Conclusions}

It is shown that in the chiral parametrization of the QCD vector
field for color $SU\left( 3\right) $ an effective color space is
defined by the hermitean color chiral field $U=m$ representing an
orbit through hypercharge $SYS^{-1}$ . The parametrization
contains an abelian field  directed along $m$ and an $U\left(
2\right)$ field $G_\mu$ which commutes with $m$ . The axial
component $\hat{\mathit{A}}_\mu $ anticommutes with $m$ and
belongs to tangential bundle of $CP^2$. Thus, the chiral
parametrization restricts the color space ascribed to gluons in
absence of quark chiral color. This parametrization is quite
different from those \cite{bolokhov}, \cite
{shabanov},\cite{walker}, where the chiral anomaly is neglected.

The $CP^2$ sector is studied. The chiral parametrization of the
vector
component $\Gamma _\mu $ involves an abelian field $C_\mu $, isospinors $%
\hat \xi ,\hat \xi ^{+}$\, as building blocks for composite fields
$B_{\mu },b_\nu ,$ $h_{\mu \nu },$ as well as an isoscalar $\omega
$. In the induced axial sector we have four components of axial
vector field in the $X-$ sector and scalars $\varphi $ and $\chi $
in axial $Y-$ sector. The abelian fields $b_\nu ,$ $h_{\mu \nu }$
are of magnetic type (in language of $ SU\left( 2\right) $). The
field strength in the vector sector $\Gamma _{\mu \nu }^{}$
describes two interacting systems. One of them is characterized be
color isospin matrix variables $\omega \hat \xi ,\omega \hat \xi
^{+}$, the second one is characterized by hypercharge like
potential $b_\mu $ with field strength $h_{\mu \nu }$. Kinetic
terms for these fields in the Yang-Mills Lagrangian are quadratic.

We hope to report applications of our parametrization of the QCD field in
future papers.

\end{document}